\documentclass{iucr}              

                   \def\href#1{\relax}\let\foo\caption
                   
\usepackage{url}

\usepackage{breakurl}
                   
\ifPDF
  \RequirePackage{hyperref}
  \PassOptionsToPackage{pdftex,bookmarksopen,bookmarksnumbered,breaklinks}{hyperref}
\fi
\let\caption\foo

     \paperprodcode{a000000}      
     \paperref{xx9999}            
     \papertype{IU}               

     \paperlang{english}          
     \journalyr{2003}
     \journalreceived{\relax}
     \journalaccepted{\relax}
     \journalonline{\relax}
\usepackage{amssymb}
\usepackage{amsthm}
\usepackage{chemformula}
\usepackage[version=3]{mhchem} 
\usepackage{physics}

\theoremstyle{definition}
\newtheorem{thm}{Theorem}
\newtheorem{dfn}[thm]{Definition}
\newtheorem{pro}[thm]{Problem}

\newcommand{\R}{\mathbb{R}}
\newcommand{\Z}{\mathbb{Z}}

\newcommand{\SE}{\mathrm{SE}}
\newcommand{\PDD}{\mathrm{PDD}}

\newcommand{\RI}{\mathrm{RI}}

\newcommand{\QT}{\mathrm{QT}}

\newcommand{\de}{\delta}

\newcommand{\ep}{\varepsilon}

\newcommand{\La}{\Lambda}

\newcommand{\vl}{\, :  \,}

\newcommand{\angstrom}{\textup{\AA}}

\newcolumntype{R}[1]{>{\raggedleft\let\newline\\\arraybackslash\hspace{0pt}}m{#1}}
\newcolumntype{L}[1]{>{\raggedright\let\newline\\\arraybackslash\hspace{0pt}}m{#1}}
\newcolumntype{C}[1]{>{\centering\let\newline\\\arraybackslash\hspace{0pt}}m{#1}}

\begin{document}                  



\title{The importance of definitions in crystallography}
\shorttitle{The importance of definitions in crystallography}


     \author[a]{Olga}{Anosova}
     \cauthor[a]{Vitaliy}{Kurlin}{vkurlin@liv.ac.uk}{}\aufn{}
     \author[b]{Marjorie}{Senechal}
     \aff[a]{Computer Science department and Materials Innovation Factory, University of Liverpool, UK}
     \aff[b]{Clark Science Center, Smith College, US}


\shortauthor{Anosova, Kurlin, and Senechal}







\maketitle                        

\begin{synopsis}
The paper proposes a rigorous definition of a periodic structure as an equivalence class of periodic sets under rigid motion, which is a composition of translations and rotations.
\end{synopsis}

\begin{abstract}
This paper was motivated by the articles ``Same or different - that is the question'' in CrystEngComm (July 2020) and ``Change to the definition of a crystal'' in the IUCr newsletter (June 2021).
Experimental approaches to crystal comparisons require rigorously defined classifications in crystallography and beyond.
Since crystal structures are determined in a rigid form, their strongest equivalence in practice is rigid motion, which is a composition of translations and rotations in 3-dimensional space.
Conventional representations based on reduced cells and standardizations 
theoretically distinguish all periodic crystals.
However, all cell-based representations are inherently discontinuous under almost any atomic displacement that can arbitrarily scale up a reduced cell.
Hence comparing millions of known structures in materials databases needs continuous distance metrics.
\end{abstract}


\section{Motivations for new definitions in crystallography}
\label{sec:intro}

Mathematical crystallography, including the  classification of lattices, unit cells, crystal classes, etc., by symmetries has a long and rich history.  
But classical mathematical crystallography, grounded largely in group theory,  was done before the computer age and needs updating in our era of massive data. 
\smallskip

This paper does not reinvent the wheel but extends the discrete concepts to a new continuous domain in the language of present-day crystallography for present-day crystallographers.
\smallskip

The entry "A crystal" appeared in the IUCr Online Dictionary of Crystallography \cite{IUCrdictionary} in 1992 and was since modified slightly.
We propose updates to fill the past gaps and meet present needs.  
The latest direct-space definition \cite{Crystal} by the Commission on Crystallographic Nomenclature (CCN) says that ``a solid is a crystal if its atoms, ions and/or molecules form, on average, a long-range ordered arrangement. In most crystals, the arrangement is a periodic array that is governed by the rules of translational symmetry.''
\smallskip

In this paper, a \emph{crystal} is a periodic crystal, so we postpone similar developments for non-periodic materials including quasicrystals and amorphous solids to future work, see \cite{senechal1996quasicrystals}.
The definition quoted above \cite{brock2021change} for a periodic crystal means that the set of all atoms is preserved under all lattice translations.
Since periodic crystals, lattices, and unit cells are often confused \cite{nespolo2015ash} or used interchangeably, we give rigorous definitions  in section~\ref{sec:confusions}.
\smallskip

The next step is to clarify which periodic crystals should be considered the \emph{same}, in order to reliably compare crystals.
Below we quote the paper ``Same or different - that is the question'' \cite{sacchi2020same}.
It is correct scientific practice ``to report measurable quantities with an error'' because all real measurements are noisy.
However, if one claims that ``... two dimensions are considered the same if their values fall within the accepted error or standard deviation'', quoted from section~2.1 in \cite{sacchi2020same}, then an axiomatic approach logically implies that all dimensions (meaning measurements of unit cell parameters in this case) should be the ``same''.
This continuum paradox \cite{hyde2011sorites} says that many small changes (indistinguishable from 0) can lead to a big overall change.  
\smallskip

For any fixed small error $\ep>0$, if we call any real number $x\in\R$ indistinguishable from (considered the same as) all numbers within an interval $[x-\ep,x+\ep]$, then $x+\ep$ is the same as all numbers from $[x+\ep,x+2\ep]$, which makes $x$ the same as any number within $[x,x+2\ep]$, similarly within $[x-2\ep,x]$ if we replace $\ep$ with $-\ep$.
Continuing this logical argument further, any number $y$ becomes indistinguishable from $x$ in $\lceil|x-y|/\ep\rceil$ steps, which is the smallest integer larger than or equal to $|x-y|/\ep$.
This argument is formalized in terms of an equivalence below.

\begin{dfn}[\emph{equivalence} relation]
\label{dfn:equivalence}
A binary relation $A\sim B$ between objects of any kind is called an \emph{equivalence} \cite{raczkowski1990equivalence} if these axioms hold:
\smallskip

\noindent
(1) \emph{reflexivity} : 
$A\sim A$, so any object $A$ is equivalent to itself;
\smallskip

\noindent
(2) \emph{symmetry} : 
for any objects $A,B$, if $A\sim B$ then $B\sim A$;
\smallskip

\noindent
(3) \emph{transitivity} : 
for any $A,B,C$, if $A\sim B$ and $B\sim C$ then $A\sim C$.
\end{dfn}

Definition~\ref{dfn:equivalence} is important because any well-defined classification into disjoint classes requires an equivalence relation.
Indeed, the \emph{equivalence class} of any object $[A]=\{B \mid B\sim A\}$ is the set of all objects $B$ equivalent to $A$.
The transitivity axiom implies that if the classes of $A,C$ share a common object $B$, these classes coincide, i.e. $[A]=[C]$.
Hence Definition~\ref{dfn:equivalence} guarantees that all equivalence classes are disjoint.
For any fixed $\ep>0$, the binary relation $x\sim y$ defined by $|x-y|\leq\ep$ on real numbers fails the transitivity axiom because $0\sim\ep\sim 2\ep$ but $0\not\sim 2\ep$.
\smallskip

If we enforce the transitivity so that $x\sim z$ if there is $y$ such that $x\sim y\sim z$, this transitive extension makes all real numbers equivalent by putting them into a single equivalence class.
Equality is an example of equivalence because any number can be written in many different forms: $0.5=\frac{1}{2}=50\%=1:2$. 
\smallskip

If the axioms of Definition~\ref{dfn:equivalence} such as the transitivity are not satisfied, the resulting classes can become overlapping and dependent on manually chosen parameters, see \cite{zwart2008surprises}. 
All relations between lattices and crystals that led to 7 crystal systems, 14 Bravais classes, and 230 space-group types are equivalences satisfying the axioms.
A space-group type is a class of space groups under \emph{isomorphism}, which is a bijection respecting the group operation, see \cite{nespolo2018crystallographic}.
\smallskip

The most important practical motivation to agree on the main equivalences between crystals is the ongoing crisis of fake data in crystallography \cite{gavezzotti2022crystallography}, which has caught attention of journalists \cite{chawla2024crystallography}.
Indeed, scientists could stop the ``paper mills'' \cite{bimler2022better} that publish hundreds of articles and thousands of crystal structures, many of which are being investigated for data integrity \cite{CSD2023blog}.
\smallskip

In November 2023, two Nature papers described the recent `big data' attempts at generating crystal structures.
The first paper \cite{merchant2023scaling} reported the GNoME database of 384+ thousand `stable' predicted structures.
The chemists found ``scant evidence for compounds that fulfill the trifecta of novelty, credibility, and utility'' \cite{cheetham2024artificial}.
\smallskip

The autonomous A-lab \cite{szymanski2023autonomous} claimed to have synthesized 43 new materials from the GNoME.
The review \cite{leeman2024challenges} concluded that ``none of the materials produced by A-lab were new: the large majority were misclassified, and a smaller number were correctly identified but already known''.
Section~\ref{sec:conclusions} will complement these conclusions by identifying thousands of duplicates in the GNoME. 
  
\section{Common confusions with cells, lattices, and crystals}
\label{sec:confusions}

In our papers \cite{widdowson2022average} and \cite{widdowson2022resolving}, we introduced a unit cell, lattice, and periodic crystal in a single definition without explaining their logical dependencies.
This approach suffices for expert mathematicians, but since many publications confuse lattices not only with crystals but also with cells, we clarify the differences here.
\smallskip

We are grateful to Massimo Nespolo for highlighting the differences between a periodic lattice and a crystal structure in \cite{nespolo2019lattice}.
Confusing these concepts led to the terms ``lattice energy'' and ``lattice defects'', which should be better called ``structural energy'' and ``structural defects''.
Since section~2 in \cite{nespolo2019lattice} defined ``the lattice of a crystal structure ... as a collection of vectors expressed as a linear combination of $n$ linearly independent vectors'', we start from the more basic concepts of a basis and a lattice without requiring a crystal structure whose definition needs the pre-requisite concept of a lattice. 

\begin{dfn}[\emph{basis} and \emph{ordered basis}]
\label{dfn:basis}
(a)
A \emph{basis} of 
$\R^n$ is an unordered set of $n$ vectors $\{\vb*{v_1},\dots,\vb*{v_n}\}$ in $\R^n$ that are \emph{linearly independent}, i.e., 
$\sum\limits_{i=1}^n t_i \vb*{v_i}=0$ if and only if $t_1=\dots=t_n=0$.
\smallskip

\noindent
(b) An \emph{ordered basis} of $\R^n$ is a basis whose vectors $\vb*{v_1},\dots,\vb*{v_n}$ are ordered.
Equivalently, any vector $\vb*{v}\in\R^n$ can be expressed as a linear combination $\sum\limits_{i=1}^n t_i \vb*{v_i}$ for unique 
$t_1,\dots,t_n\in\R$.
\end{dfn}

For example, the vectors $\vb*{v_1}=(1,0)$, $\vb*{v_2}=(0,1)$ form a basis of $\R^2$ because any vector $\vb*{v}=(x,y)\in\R^2$ is uniquely written as the linear combination $x\vb*{v_1}+y\vb*{v_2}$. 
We can write coordinates of any vector $\vb*{v}\in\R^2$ in a unique order only if $\vb*{v_1},\vb*{v_2}$ are ordered. 
\smallskip

Following our standards of introducing all concepts with an equivalence, these definitions imply that two bases are equivalent if they are equal as sets, while two ordered bases are equivalent if they contain the same vectors in the same order.
A basis is often confused with the unit cell defined by this basis.

\begin{dfn}[the \emph{unit cell} and \emph{lattice} defined by a basis]
\label{dfn:cell+lattice}
Any unordered basis $\{\vb*{v_1},\dots,\vb*{v_n}\}$ of $\R^n$ defines a \emph{unit cell}:  the parallelepiped $U(\vb*{v_1},\dots,\vb*{v_n})$ consisting of all linear combinations $\sum\limits_{i=1}^n t_i \vb*{v_i}$ with real coefficients $t_1,\dots,t_n\in[0,1)$. 
This basis also generates the  \emph{lattice} $\La(\vb*{v_1},\dots,\vb*{v_n})$ consisting of  all linear combinations $\sum\limits_{i=1}^n c_i \vb*{v_i}$ with integer coefficients $c_1,\dots,c_n\in\Z$.
\end{dfn}

\begin{figure}
\caption{\textbf{Left}: infinitely many cells generate the same square lattice.
\textbf{Right}: almost any perturbation breaks the symmetry and 
discontinuously scales a primitive cell.
}
\includegraphics[width=0.4\textwidth]{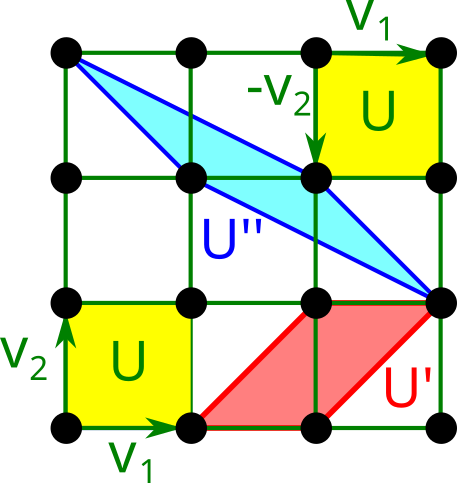}
\hspace*{2mm}
\includegraphics[width=0.55\textwidth]{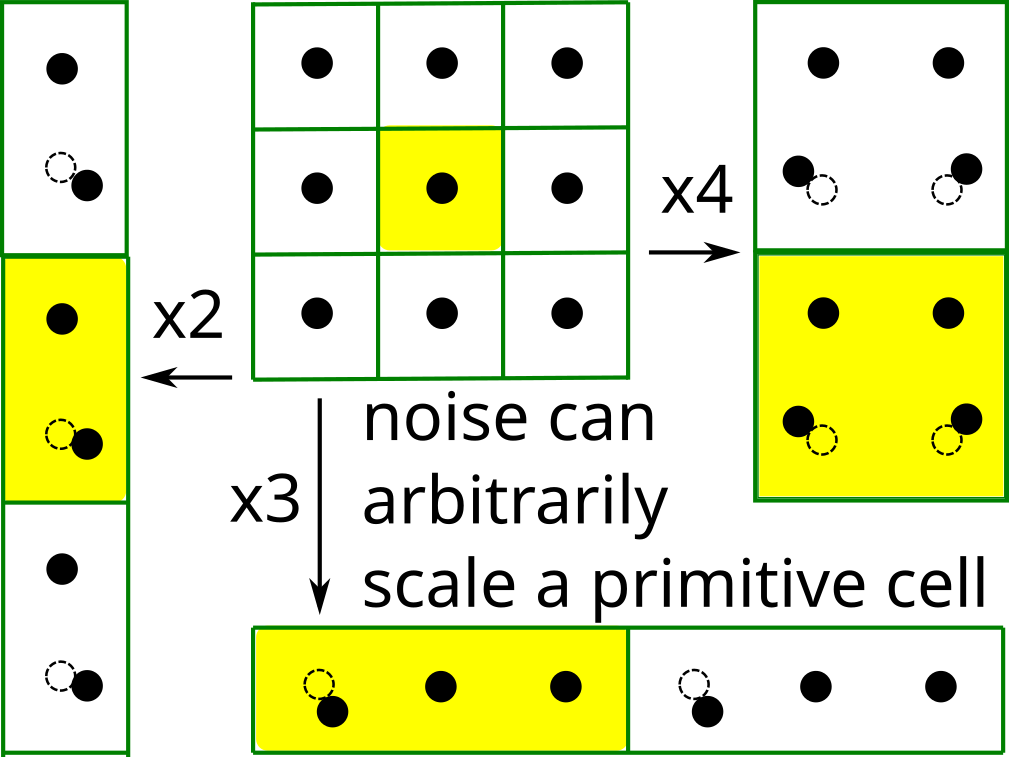}
\label{fig:ambiguity}
\end{figure}

Thus a unit cell is a "box," while a lattice is a discrete point set. 
Fig.~\ref{fig:ambiguity}~(left) shows that the square cells defined by the orthonormal bases $\{\vb*{v_1},\vb*{v_2}\}$ and $\{\vb*{v_1}, -\vb*{v_2}\}$ are both unit squares, which differ only by a choice of origin and orientation.
The square lattice has infinitely many bases $\{A\vb*{v_1},A\vb*{v_2}\}$, where
$A$ is a $2\times 2$ matrix  with integer coefficients and determinant $\pm 1$.

For the unit cell $U(\vb*{v_1},\dots,\vb*{v_n})$, we excluded the values $t_i=1$ so that all translations of $U(\vb*{v_1},\dots,\vb*{v_n})$ by vectors $\vb*{v}\in\La(\vb*{v_1},\dots,\vb*{v_n})$ tile $\R^n$ without overlaps.
The notation $U(\vb*{v_1},\dots,\vb*{v_n})$ highlights that a unit cell is defined by a basis alone.
For now, we consider unit cells (and lattices) equivalent (in the strictest possible sense) if they are equal as sets of points.
The map $\{$bases\}$\to\{$unit cells$\}$ is not invertible because a corner of a unit cell should be chosen for an origin.
Fixing one of $2^n$ corners is equivalent to choosing $n$ signs of ordered basis vectors $\pm \vb*{v_1},\dots,\pm \vb*{v_n}$.  
So we cannot uniquely identify an ordered basis from a unit cell without making one of $2^n$ choices.
\smallskip

Definition~\ref{dfn:cell+lattice} introduced a unit cell and a lattice by using only an unordered basis $\{\vb*{v_1},\dots,\vb*{v_n}\}$ of vectors.
Without these vectors, we cannot define their linear combinations.  
\smallskip

However, as soon as we need to unambiguously express a point (from a motif below) by using fractional coordinates in a basis, this basis $(\vb*{v_1},\dots,\vb*{v_n})$ should become ordered so that coordinates of any point are ordered according to the basis. 

\begin{dfn}[\emph{motif}, \emph{periodic point set}, \emph{periodic crystal}]
\label{dfn:periodic_set}
For any ordered basis $\vb*{v_1},\dots,\vb*{v_n}$ of $\R^n$, let $M\subset U(\vb*{v_1},\dots,\vb*{v_n})$ be a finite set $M$ of points. 
We call $M$  a \emph{motif}.
A \emph{periodic point set} $S=M+\La(\vb*{v_1},\dots,\vb*{v_n})$ is the set of points $\vb{p}+\vb*{v}$ for all $\vb{p}\in M$ and $\vb*{v}\in\La$.
In $\R^3$, if each point of $M$ is an atom or ion with a chemical element and charge, $S$ can be called a \emph{periodic crystal}.
\end{dfn}

In Definition~\ref{dfn:cell+lattice}, any periodic crystal has a purely geometric part, which is a periodic set of zero-sized points at all atomic centers, and the physical part of atomic attributes of these points, see the history in \cite{palgrave2021lattice}. 
Any lattice $\La$ can be considered a periodic point set whose motif $M$ consists of a single point $\vb{p}$, for example, at the origin of $\R^n$.
More general periodic crystals, even graphite, have motifs with at least two points and thus are not lattices by Definition~\ref{dfn:cell+lattice}. 
\smallskip

Any unit cell can be scaled by a positive integer factor along each basis vector to an extended cell.
This additional ambiguity is theoretically resolved by taking a \emph{primitive} cell that is a unit cell of a minimal volume.
However, Fig.~\ref{fig:ambiguity}~(right) shows that any extended cell can be made primitive by a tiny perturbation of a single atom in the initial cell.
This discontinuity was reported in 1965, see page 80 in \cite{lawton1965reduced}, and emerges even in dimension 1.
For the integer sequence $\Z$, if we shift $m$ of every $m+1$ points by a small $\ep>0$, we get the periodic sequence $\{0,1+\ep,\dots,m+\ep\}+(m+1)\Z$ whose every point is $\ep$-close to a point of $\Z$ (and vice versa) but the period $m+1$ can be arbitrarily large after perturbation.
\smallskip

\begin{figure}
\caption{For any $a>b>0$, 
let the lattices $\La,\La'\subset\R^2$ have the unit cells $U,U'$ of the rectangular forms $a\times b$, $b\times a$, respectively.
Any collection of $m\geq 2$ points with fractional coordinates $x\neq y$ in $[0,1]$ defines different motifs $M\subset U$ and $M'\subset U'$.
Then the periodic point sets $S=\La+M$, $S'=\La'+M'$ can be arbitrarily different though their CIFs differ only by swapping the lengths $a,b$.}
\includegraphics[width=\textwidth]{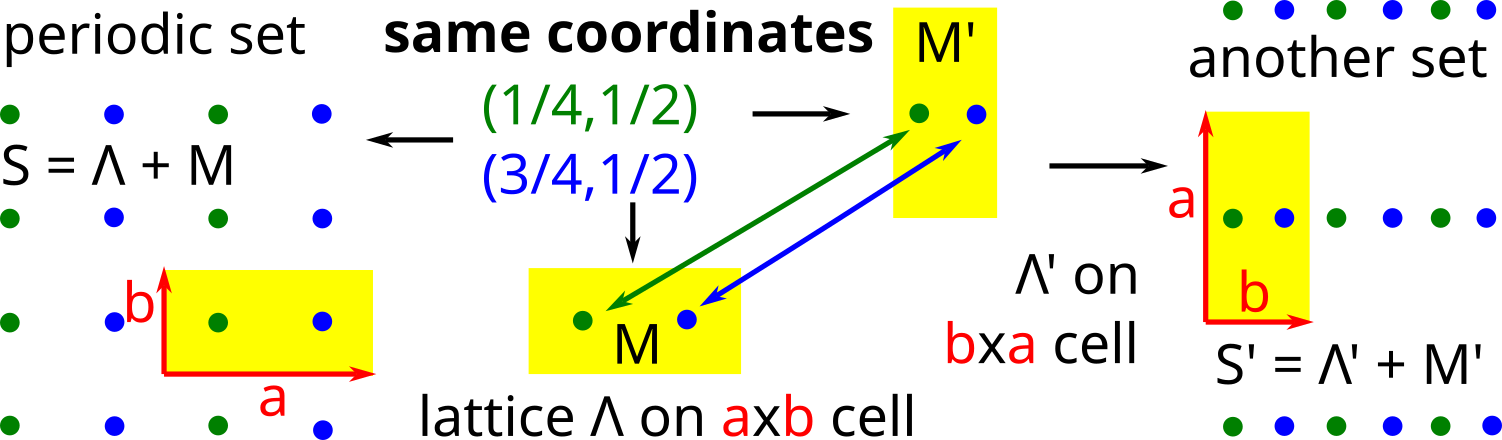}
\label{fig:reorder_rect_cell}
\end{figure}

A Crystallographic Information File (CIF) contains an ordered basis of vectors in $\vb*{v_1},\vb*{v_2},\vb*{v_3}\in\R^3$ and coordinates of each point $\vb{p}\in M$ in this basis with the atomic type of $\vb{p}$.
The ordered vectors $\vb*{v_1},\vb*{v_2},\vb*{v_3}$ can be uniquely determined from their lengths $|\vb*{v_1}|,|\vb*{v_2}|,|\vb*{v_3}|$ and angles $\angle(\vb*{v_2},\vb*{v_3})$, $\angle(\vb*{v_3},\vb*{v_1})$, $\angle(\vb*{v_1},\vb*{v_2})$.
The angles should be ordered according to their opposite vectors.
A unit cell without ordered sides (ordered basis vectors) can give rise to different periodic point sets as in Fig.~\ref{fig:reorder_rect_cell}. 

\begin{figure}
\caption{
Due to the ambiguity of Fig.~\ref{fig:reorder_rect_cell}, a unit cell $U$ with a motif $M\subset U$ can define a periodic point set only after choosing an ordered basis for $U$.
A \emph{periodic point set} is a union of lattices $\La+\vb{p}$ shifted by all $\vb{p}\in M$.
A \emph{periodic crystal} is a periodic set of atoms (points with chemical elements or other attributes).}
\includegraphics[width=\textwidth]{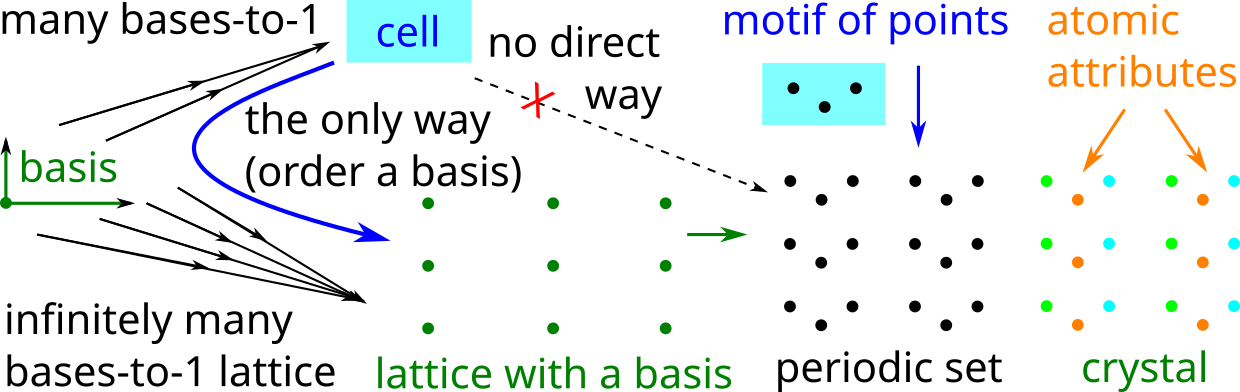}
\label{fig:basis}
\end{figure}

Ordering basis vectors by their lengths creates another discontinuity if the vectors have equal lengths because small perturbations can change their order.
Fig.~\ref{fig:basis} summarizes why an ordered basis of $\R^3$ is more convenient for defining a periodic crystal than a unit cell.
When a basis $\vb*{v_1},\dots,\vb*{v_n}$ is fixed, we use the shorter notations $U,\La$ without repeating this fixed basis.

\section{Rigorous definitions of periodic and crystal structures }
\label{sec:definitions}

In the past, many different equivalence relations between latices and crystals were studied.
One of the simplest is by chemical composition or by equality of another property such as density.
However, crystals with the same composition (say, diamond and graphite of pure carbon) or with the same density can have many different properties, so these equivalences may not suffice. 
\smallskip

Hence we are looking for a stronger equivalence that would guarantee the same physical and chemical properties according to the \emph{structure-property hypothesis} saying that a material structure should determine all its properties \cite{newnham2012structure}. 
\smallskip  

The IUCr online dictionary \cite{isostructural} contains the following entry:  ``crystals are said to be \emph{isostructural} if they have the same structure but not necessarily the same cell dimensions nor the same chemical composition, and with a 'comparable' variability in the atomic coordinates to that of the cell dimensions and chemical composition. For instance, calcite CaCO3, sodium nitrate NaNO3 and iron borate FeBO3 are isostructural''.
This phrase contains a cycle of `structural' concepts (``crystals are \emph{isostructural} if they have the same structure''), which should be resolved by defining a \emph{structure}.
\smallskip

If the keyword ``necessarily'' was omitted above, the reflexivity axiom would fail.
Any attempt to define a 'comparable' variability with a threshold $\ep>0$ for  deviations of cell sizes fails the transitivity axiom.
Indeed, by applying sufficiently many tiny deformations, we can convert any given unit cell (with an empty motif as in Definition~\ref{dfn:cell+lattice}) into any other cell, so the classification under this ``deviation'' equivalence becomes trivial.
\smallskip

The IUCr online dictionary defines the longer term \emph{crystal structure} as ``a crystal pattern consisting of atoms''.
Both \cite{pattern} and section~8.1.4 in \cite{hahn2005international} defined a \emph{crystal pattern} in different words but essentially as a periodic point set in Definition~\ref{dfn:periodic_set}, not considered under rigid equivalence.
The word \emph{pattern} as in the area of Pattern Recognition often refers not to a single object but to a class of objects under an equivalence as we propose in new Definition~\ref{dfn:structures} below.
\smallskip

Why do we need an equivalence that distinguishes between all chemical compositions and also close \emph{polymorphs} that have the same composition but different properties?
Such an equivalence is important because in the past many HIV patients suffered by unknowingly taking a more stable but less soluble polymorph of ritonavir that was accidentally manufactured instead of the required from \cite{morissette2003elucidation}.
\smallskip

On another hand, the pointwise coincidence of cells, lattices, and periodic sets from section~\ref{sec:confusions} is too strict.
Indeed, shifting the whole motif $M$ by a small vector within a fixed unit cell changes all fractional coordinates of atoms in a CIF but not the actual solid material.
We consider all equivalences and comparisons only for ideal periodic crystals and under the same ambient conditions such as room temperature and pressure. 
\smallskip

Since crystal structures are determined in a \emph{rigid form}, their strongest and practically important equivalence is rigid motion.

\begin{dfn}[rigid motion, isometry]
\label{dfn:isometry}
A \emph{rigid motion} of $\R^n$ is a composition of translations and rotations.
An \emph{isometry} of $\R^n$ is any transformation that preserves all inter-point distances.
\end{dfn}  

For an ordered basis $\vb*{v_1},\dots,\vb*{v_n}$ of $\R^n$, an \emph{orientation} can be defined as the sign of the $n\times n$ determinant with the columns $\vb*{v_1},\dots,\vb*{v_n}$.
Any orientation-preserving isometry of $\R^n$ is a rigid motion.
Any orientation-reversing isometry of $\R^n$ is a composition of one (any) mirror reflection and a rigid motion.
\smallskip

Hence isometry is a slightly weaker equivalence than rigid motion because mirror images are equivalent under isometry but not always under rigid motion.
Since mirror images can be distinguished by a (suitable chosen) sign of orientation, it almost suffices to distinguish crystals only under isometry.
\smallskip

Definition~\ref{dfn:isometry} distinguishes between \emph{isometry}, which makes sense for any metric space with no Euclidean structure, and more restrictive \emph{rigid motion} (orientation-preserving isometry).
\smallskip

The word \emph{motion} is justified by the fact that any rigid motion $f$, which excludes mirror reflections by definition, can be realized through a continuous (motion) family of isometries $f_t:\R^n\to\R^n$, where $t\in[0,1]$, $f_1=f$ and $f_0:\vb{p}\mapsto \vb{p}$ is the identity map. 
Isometry was called a symmetry operation in section~8.1.3 of \cite{hahn2005international}.
Since \emph{symmetry} has a wider meaning in science, we use the more specific concepts of \emph{rigid motion} and\emph{ isometry}.    
The comprehensive books \cite{engel2004lattice}, \cite{zhilinskii2016introduction} studied lattices through group actions.
In this language, any periodic structure from Definition~\ref{dfn:structures} is a class in the quotient of all periodic point sets under the action of the special Euclidean group $\SE(\R^n)$ of all rigid motions in $\R^n$.   

\begin{dfn}[periodic and crystal structures]
\label{dfn:structures}
A \emph{periodic structure} is an equivalence class of periodic point sets $S\subset\R^n$ under rigid motion.
A \emph{crystal structure} is an equivalence class of periodic crystals with atomic attributes under rigid motion in $\R^3$. 
\end{dfn}

Section~2 in \cite{nespolo2018crystallographic} defined a crystal structure as ``an idealized periodic pattern of atoms in three-dimensional space using the corresponding coordinates with respect to the chosen coordinate system''.
This pattern coincides with a crystal pattern \cite{pattern} from section~8.1.4 of \cite{hahn2005international} and is a single representative of a periodic structure, introduced as a class of all rigidly equivalent crystals in Definition~\ref{dfn:structures}.
\smallskip

Any explicit use of coordinates for a crystal representation as in a CIF requires choosing an ordered basis and a motif of points with fractional coordinates in this basis. 
Definition~\ref{dfn:periodic_set} called such objects periodic point sets and periodic crystals.
Shifting a motif by a fixed vector changes a description in a CIF but not the real structure considered as a class of equivalent representations
\smallskip

Then a periodic crystal in the sense of classical cell-based Definition~\ref{dfn:periodic_set} becomes one of infinitely many coordinate-based representations of a crystal structure in the sense of new Definition~\ref{dfn:structures}.
Hence crystals are defined as \emph{same} if all their atoms can be matched by rigid motion.
If there is no ideal match, any slightly different structures can be called close rather than ``the same'' because any tolerance makes the classification trivial. 
\smallskip

\begin{figure}
\caption{Feynman's first lecture in \cite{feynman1971feynman} has a table, (redrawn here in a simpler form) of 7 cubic crystals that all differ by their periodic structures geometrically as in Definition~\ref{dfn:structures} after all chemical elements are ignored. }
\includegraphics[height=36mm]{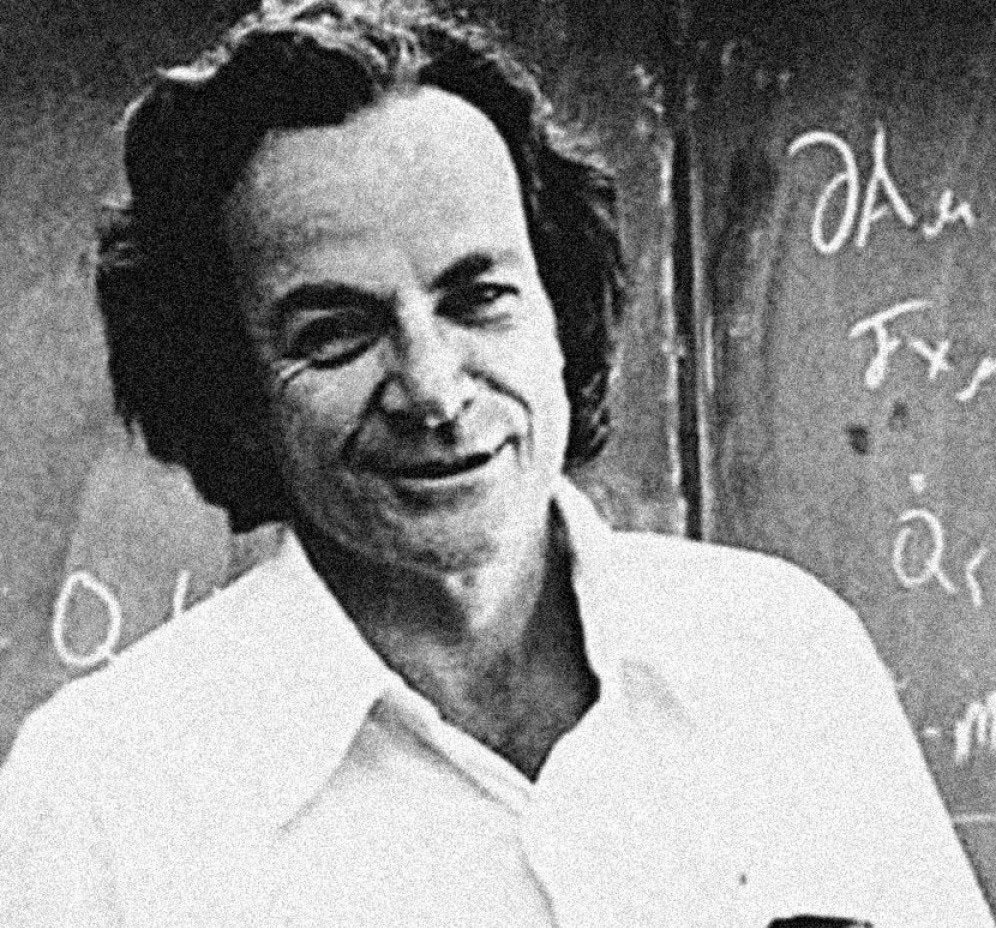}
\hspace*{1mm}
\includegraphics[height=36mm]{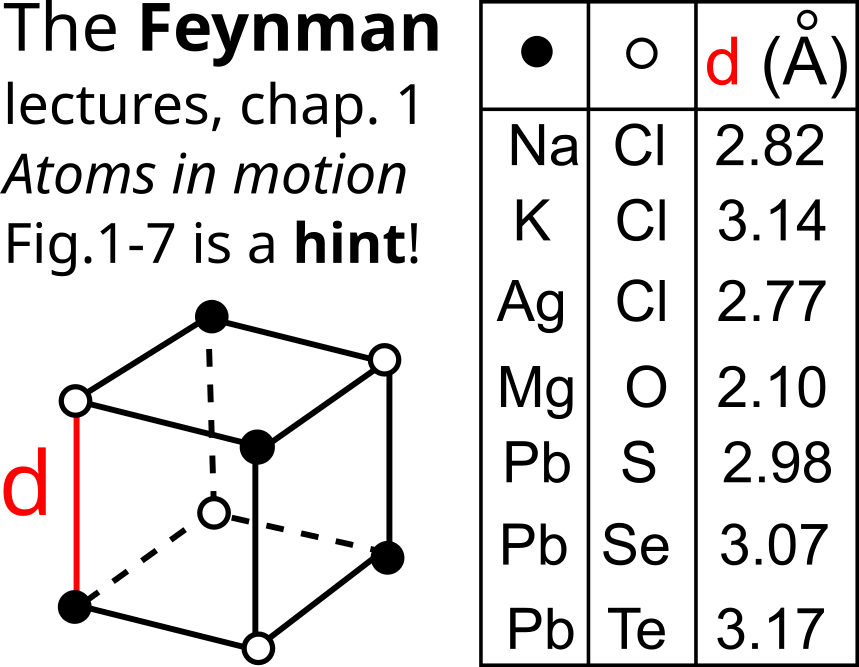}
\label{fig:Feynman}
\end{figure}

Ignoring atomic attributes maps any periodic crystal to a periodic set of points (atomic centers).
Though this projection might seem to lose all chemistry, 
Richard Feynman gave us a visual hint in his first lecture on atomic theory in Fig.~\ref{fig:Feynman} to compare crystals only by atomic centers without chemical elements. 

Despite the apparent simplicity, Definition~\ref{dfn:structures} brings up a hard problem of efficiently distinguishing periodic structures, which will be stated in section~\ref{sec:conclusions} defining a few more concepts.
A recent and almost complete solution to this problem has made Definition~\ref{dfn:structures} practically important, especially for detecting thousands of previously unknown near-duplicates in major 
databases.
Sections~\ref{sec:invariants} and~\ref{sec:metrics} will discuss how to distinguish crystal structures and continuously quantify their differences.

\section{Descriptors vs invariants under a given equivalence}
\label{sec:invariants}

Distinguishing objects under any equivalence relation from Definition~\ref{dfn:equivalence} necessarily requires the concept of an invariant.
Such a numerical property is often called a feature or descriptor without specifying an equivalence.
In the sequel, for simplicity, we use isometry as our main equivalence, denoted by $S\simeq Q$.
Extensions to rigid motion will need a sign of orientation. 

\begin{dfn}[\emph{invariant}, \emph{complete} invariant]
\label{dfn:invariant}
A function $I$ on periodic point sets is called an \emph{isometry invariant} if any isometric sets $S\simeq Q$ have $I(S)=I(Q)$ or, equivalently if $I(S)\neq I(Q)$ then $S\not\simeq Q$.
An invariant $I$ is called \emph{complete} (injective or separating) if the converse also holds: if $S\not\simeq Q$ then $I(S)\neq I(Q)$.
\end{dfn}

\begin{figure}
\caption{Any periodic set has many non-isometric subsets within boxes or balls of the same cut-off radius.
If an original basis is forgotten, it can be hard to reconstruct the initial periodic structure from its arbitrary finite subset.}
\includegraphics[width=\textwidth]{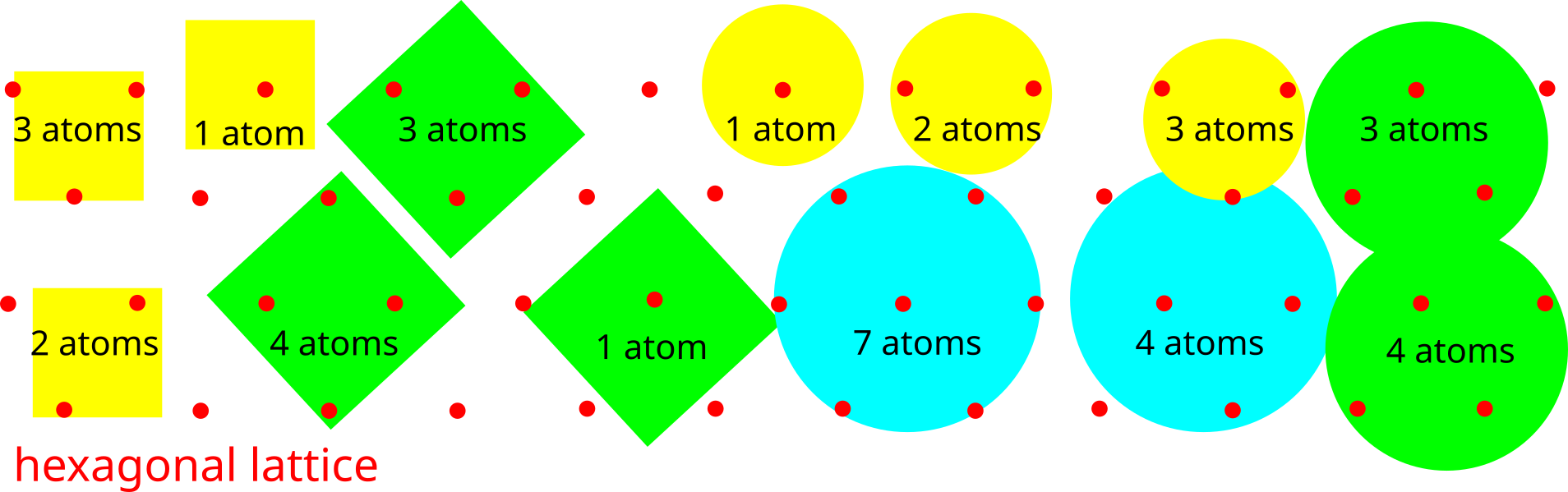}
\label{fig:finite_subsets}
\end{figure}

Though it is very tempting to reduce a periodic point set to a finite subset such as an extended motif, this reduction can lead only to many non-isometric subsets as in Fig.~\ref{fig:finite_subsets}.
Hence there is no simple way to reduce a periodic point set to a single finite subset.
Taking finite clouds around every atom in a motif can lead to a complete invariant of periodic point sets under isometry \cite{anosova2021isometry} but the continuity under perturbations needs careful justifications \cite{anosova2022algorithms}.

A simple isometry invariant of a periodic point set $S$ is the number $m$ of points within a primitive unit cell $U$ of $S$.
This invariant is weak and cannot distinguish any lattices.
A complete (injective or separating) invariant $I$ is the strongest possible in the sense that $I$ distinguishes all non-isometric sets.
\smallskip

The side-side-side (SSS) theorem from school geometry can be rephrased in terms of invariants by saying that a complete invariant of unordered three points under isometry of $\R^n$ consists of three inter-point distances up to permutations.
If all $m$ given points are ordered, the $m\times m$ matrix of their pairwise distances is complete under isometry by \cite{schoenberg1935remarks}.
\smallskip

The case of unordered points is more practical for molecules whose many atoms can be indistinguishable as in the benzene ring. 
The naive extension of distance matrices to $m$ unordered points requires $m!$ permutations, which is impractical even for small $m$.
Hence the important requirement for invariants is their computability, e.g. in polynomial time of the input size.
\smallskip

The invariance condition is the minimal requirement for a descriptor to be practically useful.
A non-invariant such as the list of fractional coordinates of all motif points $p\in M$ cannot distinguish between any periodic structures even under translation because all points of a motif $M$ can be slightly moved along the same vector within a fixed unit cell without changing the underlying periodic structure in the sense of Definition~\ref{dfn:structures}.
\smallskip

The related concept of an \emph{equivariant} means a function $E(S)$ such that any rigid motion $f$ affects $E(S)$ in a way controlled by $f$ so that $E(f(S))=T_f(E(S)$, where $T_f$ depends only on $f$ but not on $S$.
The \emph{invariance} means that $T_f$ is the identity.
\smallskip

For example, the center of mass of a finite molecule $M$ is equivariant (rigidly moves together with $M$).
But the center of mass of a motif $M$ is not equivariant for a periodic point set $S$ because a translation can push one point $p\in M$ through a side face of a unit cell $U$, so the new periodic translate of $p$ in the cell $U$ non-equivariantly changes $M$ and its center of mass.
\smallskip

Any linear combination of given point coordinates is equivariant under linear transformations, while invariants are much more restrictive and hence valuable.
Equivariants are often used for representing inter-atomic forces by vectors that should be rigidly moved with the whole structure.
Any collection of forces (one vector at every atom) can be interpreted as an ordered pair (initial structures, final structure moved by these forces).
\smallskip

Hence complete invariants suffice to describe not only static structures but also any dynamics in the space of structures. 
Mathematical crystallography developed many approaches to unambiguously identify a periodic structure under rigid motion, for example by using theoretically unique reduced cell \cite{niggli1928krystallographische}.
Then any periodic structure can have standard settings in the reduced cell \cite{parthe2013typix}.
In theory, this conventional representation is complete under rigid motion.
\smallskip 

Fig.~\ref{fig:ambiguity}~(right) shows that almost any noise can arbitrarily scale up any reduced cell.
Theorem~15 in \cite{widdowson2022average} says that this discontinuity under tiny perturbations holds even for lattices, which have motifs consisting of only one point.
\smallskip

\begin{figure}
\caption{
Non-invariants vs progressively harder requirements for isometry invariants, which will be all formalized in Problem~\ref{pro:classification}.
For periodic crystals, invariants should be computable in polynomial time of the size of a motif to be useful.}
\includegraphics[width=\textwidth]{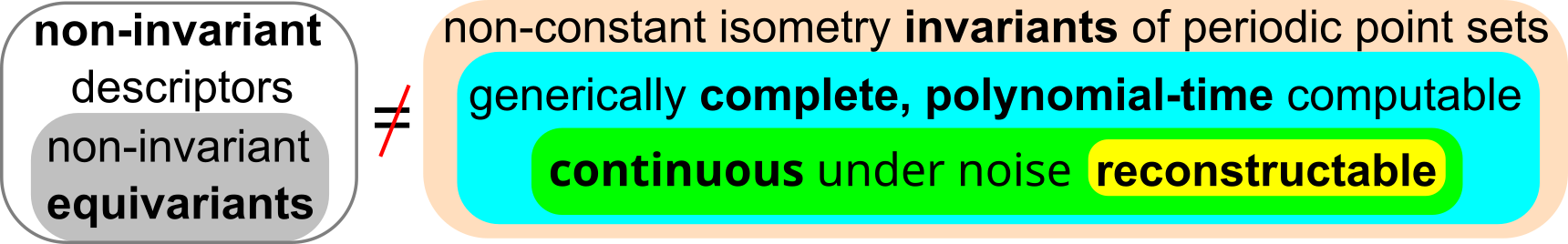}
\label{fig:invariant_types}
\end{figure}

The discontinuity of cell-based representations allows anyone to disguise a near-duplicate as a new material by making any extended cell primitive due to a slight displacement of atoms and by replacing some atoms with similar ones.
To stop potential duplicates, we need continuous invariants that can quantify any (near-)duplicates in terms of a distance metric. 
The more practically important requirements of continuity and reconstructability in Fig.~\ref{fig:invariant_types} will be formalized in section~\ref{sec:metrics}.

\section{Similarities vs distance metrics and continuity}
\label{sec:metrics}

Section~\ref{sec:invariants} justified the importance of invariants for distinguishing periodic structures.
This section formalizes the concept of continuity with respect to a distance metric.
We start from the simplest non-trivial case of 2-dimensional lattices.
\smallskip

\cite{lagrange1773recherches} classified all lattices $\La\subset\R^2$ under isometry by using the quadratic form $Q(x,y)=q_{11}x^2+2q_{12}xy+q_{22}y^2$, whose coefficients are expressed via a basis $v_1,v_2$ of a lattice $\La$ by the formulae $q_{11}=v_1\cdot v_1$, $q_{22}=v_2\cdot v_2$, $q_{12}=v_1\cdot v_2$.
The extra conditions $0<q_{11}\leq q_{22}$ and $-q_{11}\leq 2q_{12}\leq 0$ guarantee the uniqueness of the form $Q$.
The corresponding basis of $\La$ is called \emph{reduced} and is unique under isometry of $\R^2$ but not under rigid motion because the bases $v_1=(1,0)$, $v_2^{\pm}=(-\frac{a}{2},\pm b)$ for $0<a<1<b$ have the same reduced form $Q(x,y)=x^2-axy+b^2y^2$ and generate the lattices that are mirror images and not related by rigid motion of $\R^2$.
\smallskip

In a more geometric approach,
\cite{selling1874ueber} and later 
\cite{delone1934mathematical} added to any basis $\vb*{v_1},\vb*{v_2}$ of $\R^2$, the extra vector $\vb*{v_0}=-\vb*{v_1}-\vb*{v_2}$ and the restriction that all pairwise angles between these vectors are \emph{non-acute}, which means $90^\circ$ or more.
More recently, \cite{conway1992low} called such a collection $\vb*{v_0},\vb*{v_1},\vb*{v_2}$ 
an \emph{obtuse superbase}.
This name is justified by the fact that any vector $\vb*{v}\in\R^2$ can be written as $\vb*{v}=a_1\vb*{v_1}+a_2\vb*{v_2}$ for unique $a_1,a_2\in\R$ in a basis $\vb*{v_1},\vb*{v_2}$ and also as $\vb*{v}=b_0\vb*{v_0}+b_1\vb*{v_1}+b_2\vb*{v_2}$ for unique $b_0=-\dfrac{a_1+a_2}{3}$, $b_1=\dfrac{2a_1-a_2}{3}$, $b_2=\dfrac{2a_2-a_1}{3}$ so that $b_0+b_1+b_2=0$.
\smallskip

\begin{figure}
\caption{
Any lattice $\La\subset\R^2$ has an obtuse superbase of basis vectors $\vb*{v_1},\vb*{v_2}$ with $\vb*{v_0}=-\vb*{v_1}-\vb*{v_2}$ and $\vb*{v_i}\cdot \vb*{v_j}\leq 0$ for distinct $i,j\in\{0,1,2\}$, which is unique under isometry, but not under rigid motion (for a rectangular lattice on the right). }
\includegraphics[width=\textwidth]{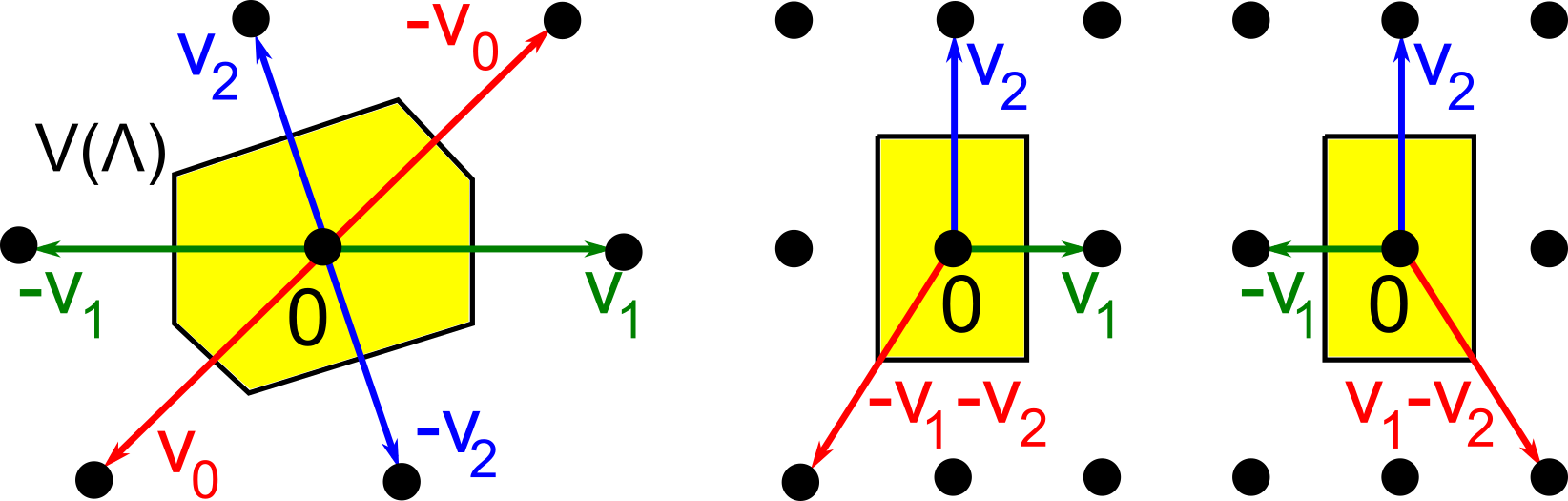}
\label{fig:obtuse_superbases}
\end{figure}

While any lattice in $\R^2$ has infinitely many non-isometric bases, see Fig.~\ref{fig:ambiguity}~(left), its obtuse superbase is unique up to isometry.
Indeed, any non-rectangular lattice $\La\subset\R^2$ has only two opposite superbases $\pm\{\vb*{v_0},\vb*{v_1},\vb*{v_2}\}$, which are related by the 2-fold rotation around $0\in\R^2$, and whose all six vectors are orthogonal to the boundary of the hexagonal Voronoi domain $V(\La)=\{\vb{p}\in\R^2\vl |\vb{p}|\leq |\vb*{v}| \text{ for } \vb*{v}\in\La-\{0\}\}$ in Fig.~\ref{fig:obtuse_superbases}~(left), see \cite{voronoi1908nouvelles}.
All obtuse superbases of a rectangular lattice are related by reflections and are not unique under rigid motion.
Fig.~\ref{fig:obtuse_superbases}~(right) shows two obtuse superbases (mirror images) for $\vb*{v_1}=(a,0)$, $\vb*{v_2}=(0,b)$, $\vb*{v_0}=(-a,-b)$. 

\begin{dfn}[\emph{root invariant} $\RI(\La)$ of a lattice $\La\subset\R^2$]
\label{dfn:root_invariant}
Let a lattice $\La\subset\R^2$ have an obtuse superbase $\vb*{v_0},\vb*{v_1},\vb*{v_2}$ so that $\vb*{v_1},\vb*{v_2}$ generate $\La$, $\vb*{v_0}+\vb*{v_1}+\vb*{v_2}=0$, and $\vb*{v_i}\cdot \vb*{v_j}\leq 0$ for all distinct $i,j\in\{0,1,2\}$.
Write the \emph{root products} $r_{ij}=\sqrt{-\vb*{v_i}\cdot \vb*{v_j}}$ in increasing order $0\leq r_{12}\leq r_{01}\leq r_{02}$, which might re-order the vectors $\vb*{v_0},\vb*{v_1},\vb*{v_2}$ without changing $\La$.
The \emph{root invariant} is the ordered triple $\RI(\La)=(r_{12},r_{01},r_{02})$, where only $r_{12}$ can be 0.
\end{dfn}

Theorem~4.2 in \cite{kurlin2022mathematics} proved that $\RI(\La)$ is a complete invariant of all lattices $\La\subset\R^2$ under isometry, also under rigid motion after enriching $\RI(\La)$ with a sign of orientation.
The key advantage of $\RI(\La)$ in comparison with a reduced basis is the continuity under perturbations.
In \cite{kurlin2022mathematics}, Fig.~4 explains the discontinuity of reduced bases, while Theorems 7.5 and 7.7 prove the bi-continuity of the root invariant $\RI(\La)$. 
\smallskip

\begin{figure}
\caption{
For each crystal in the CSD with a given basis $\vb*{v_1},\vb*{v_2},\vb*{v_3}$, we took three lattices generated by the bases $(\vb*{v_1},\vb*{v_2})$, $(\vb*{v_2},\vb*{v_3})$, $(\vb*{v_3},\vb*{v_1})$.
The resulting 2.6+ million 2D lattices populate a triangle continuously expanding the tree of Bravais classes.
The color indicates a logarithmically scaled 
number of lattices whose invariants are close to $(x,y)$, see the earlier version in Fig.~9 of \cite{bright2023geographic}. }
\includegraphics[width=\textwidth]{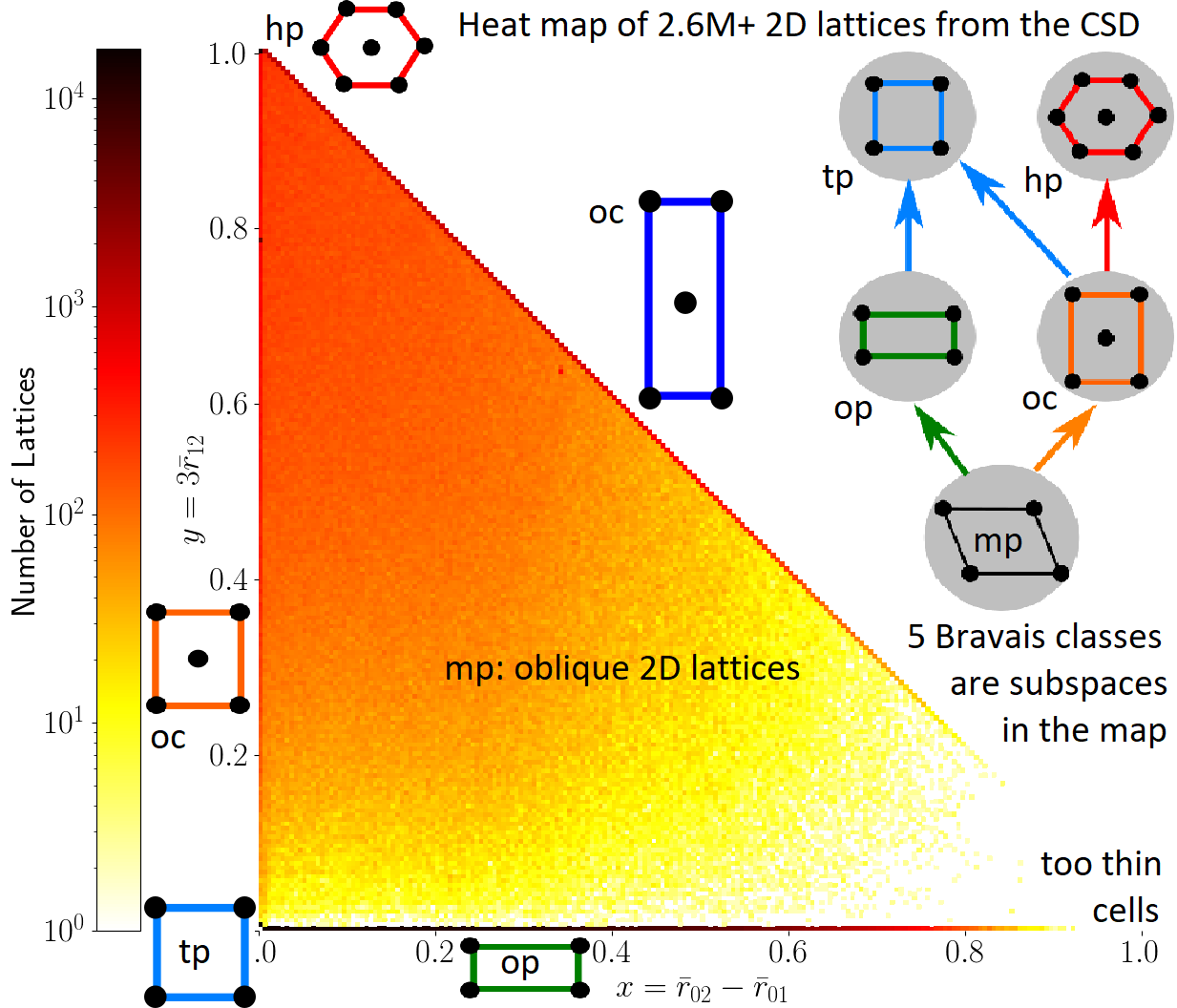}
\label{fig:QT_CSD}
\end{figure}

Fig.~\ref{fig:QT_CSD} visualizes the continuous space of all 2-dimensional lattices under isometry composed (for simplicity) with uniform scaling, which maps each root product to $\bar r_{ij}=\dfrac{r_{ij}}{r_{12}+r_{01}+r_{02}}$.
Since $\bar r_{12}+\bar r_{01}+\bar r_{02}=1$, we can use only two independent coordinates $x=\bar r_{02}-\bar r_{01}$ and $y=3\bar r_{12}$, which define the \emph{quotient triangle} $\QT=\{x+y\leq 1, x\in[0,1), y\in[0,1]\}$.
Any rectangular lattice $\La(a,b)$ with an obtuse superbase $\vb*{v_1}=(a,0)$, $\vb*{v_2}=(0,b)$, $\vb*{v_0}=(-a,-b)$ for $a\leq b$ has $\RI(a,b)=(0,a,b)$ and $(x,y)=(\frac{b-a}{b+a},0)$.
All square lattices with $a=b$ are represented by the origin $(x,y)=(0,0)$.
The point $(1,0)$ is excluded as a limit case of lattices with infinitely thin and long cells.
\smallskip

In summary, all classes of 2-dimensional lattices under isometry and uniform scaling are in a 1-1 bi-continuous correspondence with all points in the quotient triangle $\QT$.
The Bravais classes of square and hexagonal lattices are the points $(0,0)$ and $(0,1)$, respectively.
The Bravais class of centered rectangular lattices consists of two boundary edges (without endpoints): the hypotenuse $x+y=1$ and vertical side $x=0$, $y\in(0,1)$.
\smallskip

Any continuous path in $\QT$ is realized as a continuous deformation of lattices.
For example, the unit square lattice $\La_0$ with the obtuse superbase $(3,0)$, $(0,3)$, $(-3,-3)$, and $\RI(\La_0)=(0,3,3)$ can be continuously deformed into the hexagonal lattice $\La_1$ with the obtuse superbase $(2\sqrt{2},0)$, $(-\sqrt{2},\pm\sqrt{6})$, and $\RI(\La_1)=(2,2,2)$ along the vertical side $x=0$, $y\in(0,1)$ through the lattices $\La_y$ with $\RI(\La_y)=(2y,3-y,3-y)$ and the bases $\vb*{v_1}=(l,0)$ and $\vb*{v_2}=(-4y^2/l,\sqrt{l^2-16y^4/l^2})$, where $l=\sqrt{5y^2-6y+9}$, and $y$ continuously moves from $0$ to $1$.
\smallskip

Fig.~\ref{fig:QT_CSD} contrasts the discrete tree of five Bravais classes of 2-dimensional lattices with the continuous map on the quotient triangle $\QT$.
Although every orthorhombic crystal from the CSD is represented by three rectangular lattices (on three pairs of reduced basis vectors), about 45\% of all resulting lattices are oblique and continuously fill the interior of $\QT$ apart from the sparse corner close to $(1,0)$, where lattices have very thin and long primitive unit cell.
All non-generic lattices occupy lower-dimensional subspaces in the continuous space of lattices.
\smallskip

One can define many continuous distances between points in the quotient triangle $\QT$ in Fig.~\ref{fig:QT_CSD}, hence between classes of 2D lattices under isometry and uniform scaling.
In \cite{kurlin2022mathematics}, section~5 gave closed-form expressions for metrics between root invariants and section~6 quantified deviations from symmetry by continuous chiral distances, see \cite{bright2023continuous}.
\smallskip
 
Any lattice in $\R^3$ has an obtuse superbase, which is unique under isometry only for generic lattices whose Voronoi domain is a truncated octahedron. 
Lemmas 4.1-4.5 in \cite{kurlin2022complete} explicitly described all non-isometric obtuse superbases for the five Voronoi types of 3-dimensional lattices. 
These results led to a complete root invariant of lattices under isometry in $\R^3$ in \cite{kurlin2022complete}.
The root invariant of a 3D lattice requires complicated continuous distances satisfying the metric axioms in Definition~\ref{dfn:metric} below and will appear in a forthcoming work. 
\smallskip

The even more general case of periodic point sets needs a metric satisfying the axioms below.
This metric is a distance between two objects, not a numerical property of a single object.

\begin{dfn}[distance metric]
\label{dfn:metric}
For any objects under an equivalence relation $A\sim B$ from Definition~\ref{dfn:equivalence}, a distance metric $d(A,B)$ is a function satisfying these axioms:
\smallskip

\noindent
(1) \emph{coincidence}: 
$d(A,B)=0$ if and only if $A\sim B$;
\smallskip

\noindent
(2) \emph{symmetry}: 
$d(A,B)=d(B,A)$ for any objects $A,B$;
\smallskip

\noindent
(3) $\triangle$ \emph{inequality}: 
$d(A,B)+d(B,C)\geq d(A,C)$ for any $A,B,C$. 
\end{dfn}

The positivity property $d(A,B)\geq 0$ follows the axioms above.
A metric is needed to formalize the continuity of invariants in Problem~\ref{pro:classification} below. 
While classical crystallography theoretically achieved the completeness of cell-based invariants, Problem~\ref{pro:classification} asks for more practically important invariants that have no discontinuities at boundaries of 230 (or any other number of) classes in the fully connected crystal universe.

\begin{pro}[isometry classification of periodic structures]
\label{pro:classification}
Find a function $I$ on all periodic point sets $S\subset\R^n$ satisfying the following practically important conditions:
\smallskip

\noindent
(a) \emph{invariance} : 
if $S\simeq Q$ are isometric, then $I(S)=I(Q)$;
\smallskip

\noindent
(b) \emph{completeness} : 
if $I(S)=I(Q)$, then $S\simeq Q$ are isometric;
\smallskip

\noindent
(c) \emph{continuity} :
there is a metric $d$ satisfying the axioms of Definition~\ref{dfn:metric} under isometry and the $\ep-\de$ continuity below: 
for any $\ep>0$ and a periodic point set $S$, there exist $C$ and $\de>0$ such that if $Q$ is obtained by perturbing any point of $S$ up to $\de$ in Euclidean distance, then $d(I(S),I(Q))\leq C\ep$;
\smallskip

\noindent
(d) \emph{reconstructability} :
any periodic point set $S\subset\R^n$ can be reconstructed (uniquely up to isometry) from its invariant $I(S)$;
\smallskip

\noindent
(e) \emph{computability} :
the invariant $I$, metric $d$, and reconstruction of $S\subset\R^n$ can be obtained in polynomial time of the motif size from a suitably reduced basis of $S$ and motif points in this basis.
\end{pro}

Due to the coincidence axiom of a metric in Definition~\ref{dfn:metric}, the equality $I(S)=I(Q)$ in the completeness~\ref{pro:classification}(b) is best checked as $d(I(S),I(Q))=0$.
If computability~\ref{pro:classification}(e) is missed, one impractical invariant $I(S)$ satisfying all other conditions can be defined as the isometry class of all (infinitely many) periodic point sets isometric to $S$.
We assume that a periodic point set $S$ is given with a reduced basis such as Niggli's basis in $\R^3$ or Minkowski basis in a higher dimension $n$ since lattice reductions can be slow for $n>3$, see 
 \cite{nguyen2009low}.
\smallskip

The $\ep-\de$ continuity in condition~\ref{pro:classification}(c) is a classical but weak version of continuity.
The stronger \emph{Lipschitz} continuity says that $C$ and $\de$ are independent of $S$ and $\ep$, so if $Q$ is $\ep$-close to $S$, then $d(I(S),I(Q))\leq C\ep$, where a constant $\de$ was absorbed by $C\ep$.
\smallskip

For 2D lattices $\La$, Theorem~7.5 in \cite{kurlin2022mathematics} proved the intermediate H\"older continuity saying that if the coordinates of the basis vectors of $\La$ are perturbed up to $\ep$, the root invariant $\RI(\La)$ changes up to $\sqrt{6l\ep}$ in the Euclidean metric, where $l$ is the maximum length of given basis vectors of $\La$. 
\smallskip

The stronger Lipschitz continuity (without the factor $\sqrt{l}$) seems unrealistic for lattices because the rectangular lattices with the $\ep$-close bases $(l,0),(0,\ep)$ and $(l,0),(0,2\ep)$ can substantially differ even by unit cell areas $l\ep$ and $2l\ep$ whose difference $l\ep$ can be arbitrarily large if 
$l$ has no upper bound.  
\smallskip

\begin{figure}
\caption{To explain the structure-property relations, a crystal structure $S$ with infinitely many representations under isometry should be bijectively mapped by a complete and continuous invariant $I$ to the space of invariants so that any image $I(S)$ can be efficiently inverted back to a representative crystal $S\subset\R^3$.}
\includegraphics[width=\textwidth]{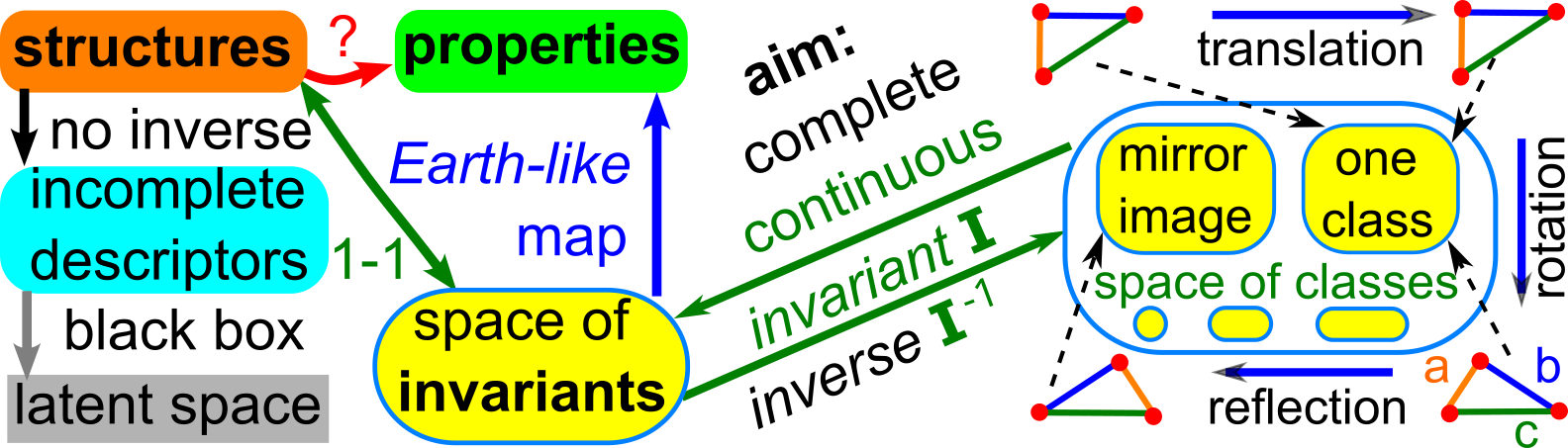}
\label{fig:structure-property}
\end{figure}

Fig.~\ref{fig:structure-property} visualizes the advantages of invariants that satisfy all the conditions of  Problem~\ref{pro:classification}.
In the past, incomplete, discontinuous or non-invariant descriptors mapped periodic crystals to latent spaces (image spaces of descriptor functions).
\smallskip

The \emph{non-invariance} (existence of false negatives) means that the same crystal structure maps to different points, which makes the problem of distinguishing structures even harder.
The \emph{incompleteness} (existence of false positives) means that non-isometric structures map to the same point, which leaves no chance to reconstruct a correct crystal.
The \emph{discontinuity} under tiny atomic displacements means that near-duplicates can appear very distant in the latent space.
\smallskip

All the conditions of Problem~\ref{pro:classification} guarantee that a required invariant $I$ is a bijective and continuous map from the space of crystal structures to the space of invariant values.
The inverse map $I^{-1}$ reconstructs any periodic point set $S$ from
$I(S)$. 

\section{Conclusions: the practical importance of definitions}
\label{sec:conclusions}

This section summarizes the progress in developing invariants that satisfy the conditions of Problem~\ref{pro:classification}.
The root invariant from Definition~\ref{dfn:root_invariant} satisfies all conditions of Problem~\ref{pro:classification} for all 2-dimensional lattices even with the stronger H\"older continuity (instead of the weaker $\ep-\de$ continuity) under rigid motion, which is stronger than isometry.
For 3-dimensional lattices, \cite{kurlin2022complete} defined a complete isometry invariant whose continuity under perturbations is being finalized.
\smallskip

The past approaches defined metrics between lattices that allowed only slow or approximate computations. 
Some of these theoretical metrics were proved to be continuous for isometry classes of lattices in any dimension \cite{mosca2020voronoi}.
\smallskip

In \cite{widdowson2022resolving} for arbitrary periodic point sets $S$ in $\R^n$, Definition~3.3 defined the \emph{Pointwise Distance Distribution} $\PDD(S;k)$, where $k$ is the number of neighbors taken into account for any point in a motif.
Theorem 4.3 proved the Lipschitz continuity saying that perturbing any atom up to $\ep$ changes $\PDD(S;k)$ only up to $2\ep$ in a suitable metric.
Theorem 4.4 proved that $\PDD(S;k)$ is \emph{generically complete} in the sense that almost any periodic structure $S\subset\R^n$ (outside singular subspaces of measure 0) can be reconstructed from a lattice of $S$ and $\PDD(S;k)$ with an explicit upper bound on $k$ 
depending on a given unit cell and motif of $S$.
Hence $\PDD$ can be considered a DNA-style code that uniquely identifies almost any real periodic crystal.
$\PDD$ is stronger for periodic crystals than DNA, which allows identical twins (about 0.3\% among humans) with indistinguishable DNAs, see \cite{osterman2022births}.
\smallskip

In practice, $\PDD(S;100)$ distinguished all (more than 660 thousand) different periodic crystals in the Cambridge Structural Database (CSD) through more than 200 billion pairwise comparisons, which were completed within two days on a modest desktop.
Section 6 in \cite{widdowson2022resolving} lists several pairs that turned out to be near-duplicate CIFs, where all numbers (unit cell parameters and fractional coordinates) were identical almost to the last decimal place but one atom was replaced with a different one, e.g. Cd with Mn in the pair JEPLIA vs HIFCAB.
The integrity office of the Cambridge Crystallographic Data Centre and all other crystallographers who looked at these previously unknown near-duplicates agreed that such an atomic replacement should more substantially perturb the geometry of atomic centers, so five journals are investigating the data integrity of the underlying publications.
\smallskip

A forthcoming paper will extend $\PDD$ invariants to distinguish all known pairs of \emph{homometric} crystals that (by definition) have the same (infinite) list of all interatomic distances.
We conjecture that the extended invariants are theoretically complete for all periodic point sets under isometry in any Euclidean $\R^n$. 
\smallskip

The comparisons above use only geometry of atomic centers without chemical elements.
After excluding the unrealistic duplicates found in the CSD, the $\PDD$ invariants mapped all non-isometric crystal structures to non-isometric periodic structures, where each atom is replaced with a zero-sized point.
\smallskip

\begin{figure}
\caption{The \emph{Crystal Isometry Principle} says that all atomic types in real periodic crystals can be reconstructed from the geometry of atomic centers given with enough precision, first stated in section~6 of \cite{widdowson2022average} and inspired by Feynman's visual hint in Fig.~\ref{fig:Feynman}, see Fig.~1-7 in \cite{feynman1971feynman}.}
\includegraphics[width=\textwidth]{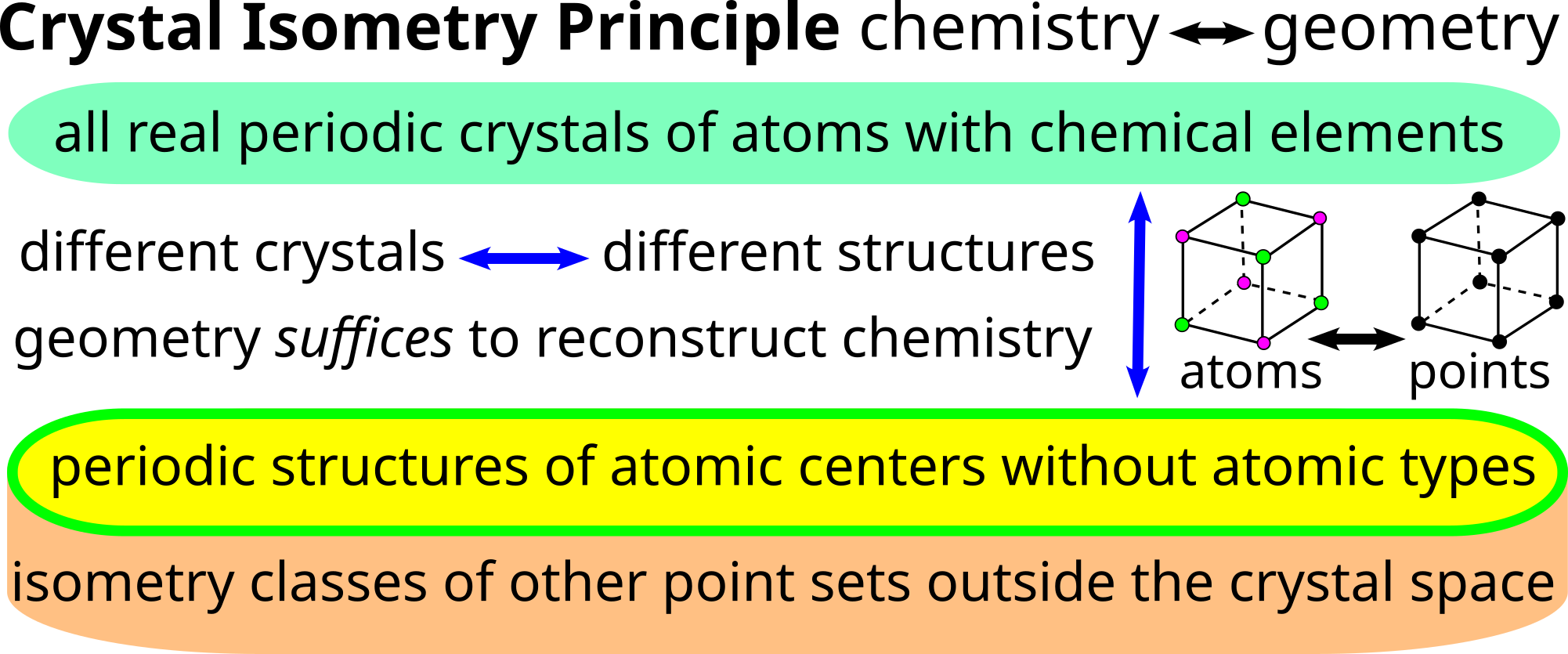}
\label{fig:CRISP}
\end{figure}

Since this map is injective, the more important conclusion is the \emph{Crystal Isometry Principle} (CRISP) saying that any real periodic structure has a unique location in a common \emph{Crystal Isometry Space} of all periodic structures (isometry classes of periodic point sets) independent of symmetry, see Fig.~\ref{fig:CRISP}.
\smallskip

Hence, in principle, all atomic types in a real periodic crystal can be reconstructed from a sufficiently precise geometry of their atomic centers.
The Eureka moment for this insight happened in May 2021 when the second author was reading Richard Feynman's first lecture ``Atoms and motion'', see Fig.~\ref{fig:Feynman} with the table of seven cubic crystals whose chemistry can be reconstructed from the only geometric parameter $d$ (smallest inter-atomic distance) known to two decimal places. 

The Crystal Isometry Principle does not claim that any periodic point set gives rise to a real periodic crystal because inter-atomic distances cannot be arbitrary.
However, any newly discovered periodic crystal will appear in the same continuous universe, where all known crystals are already visible.
Fig.~\ref{fig:QT_CSD} showed a map of 2-dimensional lattices under isometry and uniform scaling.
Continuous maps of the CSD and other databases in invariant coordinates were presented at the IUCr congress, see \cite{kurlin2023crystal}, and will be discussed in future work.
\smallskip

While the realizability of root invariants by lattices in dimension 2 and 3 has been established in \cite{kurlin2022mathematics} and \cite{kurlin2022complete}, we keep working on the harder problem of realizability of PDD invariants.
The implemented application of PDD is the ultra-fast detection of (near-)duplicates in structural databases.
Final sections in \cite{widdowson2022average} and \cite{widdowson2022resolving} reported over a dozen such pairs in the CSD.
Another forthcoming work will report less obvious (near-)duplicates in the CSD and many more duplicates in the Crystallography Open Database (COD), Inorganic Crystal Structure Database (ICSD), Materials Project, and others.
\smallskip

The most important practical impact of CRISP is the scientific barrier for ``paper mills'' and ``duplicate generators'' that can output thousands and even millions of `predicted' and sometimes `synthesized' materials by disguising known structures as new by tiny perturbations of cell parameters and atomic coordinates (structure factors or other experimental data if needed) to scale up a primitive cell, and finally by changing some non-standard chemical elements to their suitable neighbors in the periodic table.
Google's example below shows that even big numbers cannot mask (near-)duplicates that we can filter out by numbers in given CIFs even before computing invariants.
\smallskip

The paper finishes by describing embarrassing coincidences in Google's GNoME database of 384,398 `stable' structures in \cite{google2023} predicted by expensive DFT optimization \cite{mardirossian2017thirty}.
The following \emph{crystal bug test} can substantially reduce further invariant computations for such a large database.
Ordering all CIFs by the unit cell volume detected many thousand pairs of CIFs  in the GNoME that have identical volumes to all (eight) decimal places (digits).
\smallskip

Other colleagues found some duplicates after ordering all CIFs by file sizes in bytes but filtering by the unit cell volume is more justified.  
Further filtering by six parameters (three lengths and three angles) of a unit cell found 30+ thousand CIFs with identical unit cells, again with all given digits.
\smallskip

\begin{table}
\caption{Coincidences across all CIFs in the GNoME database of 384,398 publicly available CIFs \cite{google2023}.
The first column shows the sizes of the found groups whose CIFs are (near-)duplicates.
Columns 2,3,4,5 count fully identical (symbol-by-symbol) CIFs, the CIFs whose all numbers (unit cell parameters and fractional coordinates) coincide with all digits (at least 6), then CIFs whose all numbers coincide up to 4 and 2 digits, respectively.
The last row counts the total number of the involved  CIFs.
The largest groups are listed in Table~\ref{tab:GNoME_groups}.
}
\label{tab:GNoME_coincidences}
\begin{tabular}{L{12mm}|L{16mm}|L{14mm}|L{12mm}|L{12mm}} 
group size $=$ \#CIFs & groups of identical CIFs  & all numbers coincide & rounding to 4 digits & rounding to 2 digits \\
 \hline
 10 & 0 & 0 & 0 & 1 \\ 
 9   & 0 & 1 & 1 & 0 \\ 
 7   & 0 & 1 & 1 & 2 \\
 6   & 0 & 2 & 2 & 4 \\
 5   & 0 & 2 & 3 & 18 \\
 4   & 1 & 8 & 12 & 92 \\
 3   & 43 & 72 & 96 & 670 \\
 2   & 1,089 & 1,481 & 1,932 & 7,856 \\
\hline
 all CIFs & 2,311 & 3,248 & 4,243 & 18,228
\end{tabular}
\end{table}

Table~\ref{tab:GNoME_coincidences} summarizes more hard-to-explain 
coincidences. 
The CIFs with GNoME's ids 4135ff7bc7, 6370e8cf86, c6afea2d8e, e1ea534c2c are identical texts (symbol-by-symbol).
The supplementary materials (available by request) contain an Excel table listing more than a thousand pairs of identical CIFs.
If chemical elements are ignored, the GNoME has 1,481 pairs of CIFs with all equal numbers (unit cell parameters and fractional coordinates).
If we round all numbers to 4 and 2 decimal places for the precision of $10^{-4}\angstrom$ and $10^{-2}\angstrom$, respectively, the last two columns in Table~\ref{tab:GNoME_coincidences} show many more groups of CIFs that become numerically identical to each other.
Table~\ref{tab:GNoME_groups} shows chemical compositions for the three largest groups of CIFs.
\smallskip

The first part of Table~\ref{tab:GNoME_groups} says that the GNoME contains a group of 9 CIFs, where all numbers are equal (with all decimal places) but chemical compositions differ by one or two atoms.
For example, Dy, Y, Ho, and Tb are often swapped.
If all numbers are rounded to two digits, one more CIF (a18d30a9fc) joins the group of duplicates, where Ru is replaced with Re.
So comparisons by unit cell parameters and fractional coordinates can help to filter out obvious (near-)duplicates even in big data.
\smallskip

In conclusion, this paper clarified the concept of a periodic crystal in terms of an ordered basis whose re-ordering creates ambiguity or discontinuity in Fig.~\ref{fig:reorder_rect_cell}. Definitions~\ref{dfn:basis}, \ref{dfn:cell+lattice}, and \ref{dfn:periodic_set} are visually summarized in Fig.~\ref{fig:basis}. 
Rigid motion (or slightly weaker isometry) is motivated as the strongest equivalence between crystals whose structures are determined in a rigid form.
The practical importance of distinguishing near-duplicates in major structural databases requires us to define a periodic (crystal) structure as an \emph{equivalence class under rigid motion}.
Any deviations from an ideal rigid matching should be continuously quantified in terms of a distance metric satisfying all axioms and at least the classical $\ep-\de$ form of continuity.
\smallskip

\begin{table}
\caption{The largest groups of (near-)duplicates from Table~\ref{tab:GNoME_coincidences} in the GNoME database.}
\label{tab:GNoME_groups}
\begin{tabular}{l|l|R{17mm}|R{21mm}} 
GNoME id & chemical formula & all numbers in CIFs coincide & numbers coincide up to 2 digits \\
 \hline
082738d51d & \ce{Dy1 Y6 Ho13 Cd6 Ru2} & in a group of 9 & in a group of 10\\
1fba8c028f & \ce{Dy_2 Y_4 Ho_{14} Cd_6 Ru_2} & 9 & 10\\
39fe92e2ee & \ce{Tb_2 Y_4 Ho_{14} Cd_6 Ru_2} & 9 & 10\\
6d47ae3d9f & \ce{Tb_3 Y_3 Ho_{14} Cd_6 Ru_2} & 9 & 10\\
703ed1d823 & \ce{Tb_6 Ho_{14} Cd_6 Ru_2} & 9 & 10\\
78fcd9d814 & \ce{Tb_1 Y_5 Ho_{14} Cd_6 Ru_2} & 9 &10 \\
976f8cb279 & \ce{Y_6 Ho_{14} Cd_6 Ru_2} & 9 & 10\\
a30e9d8c9b & \ce{Tb_5 Y_1 Ho_{14} Cd_6 Ru_2} & 9 & 10\\
b8c0e953e2 & \ce{Tb_4 Y_2 Ho_{14} Cd_6 Ru_2} & 9 & 10\\
a18d30a9fc & \ce{Tb_6 Ho_{14} Cd_6 Re_2} & in a group of 1 & 10 \\
\hline
06eb60e958 & \ce{Li_2 Tb_2 Ho_4 Hg_8} & in a group of 7 & in a group of 7\\
9762be0ec6 & \ce{Li_2 Tb_2 Dy_4 Hg_8} &7 &7\\
ab336b54ee & \ce{Li_2 Tb_2 Er_4 Hg_8} &7 &7\\
aed8780f34 & \ce{Na_2 Tb_2 Lu_4 Hg_8} &7 &7\\
c2236e05de & \ce{Na_2 Tb_2 Dy_4 Hg_8} &7 &7\\
ca1d14568f & \ce{Na_2 Tb_2 Tm_4 Hg_8} &7 &7\\
d9eab4539b & \ce{Li_2 Tb_2 Y_4 Hg_8} &7 &7\\
\hline
02c4cb55a6 & \ce{Tb5 Dy15 Cd6 Ru2} & in a group of 6& in a group of 7\\
0affe9c149 & \ce{Tb2 Dy18 Cd6 Ru2} & 6 & 7\\
100cfdfdef & \ce{Tb3 Dy17 Cd6 Ru2} & 6 & 7\\
877c190805 & \ce{Tb4 Dy16 Cd6 Ru2} & 6 & 7\\
9ce48821cb & \ce{Dy20 Cd6 Ru2} & 6 & 7\\
b9e4b78276 & \ce{Tb1 Dy19 Cd6 Ru2} & 6 & 7\\
cf7af6f79f & \ce{Dy9 Y6 Ho5 Cd6 Ru2} & in a group of 1 & 7\\
\end{tabular}
\end{table}

As a visual summary, Fig.~\ref{fig:invariant_types} highlights the importance of invariants vs non-invariant descriptors. Fig.~\ref{fig:finite_subsets} explains that similarities based on single (hence non-invariant) finite subsets are hard to justify for periodic structures.
In the past, crystallography developed conventional representations based on reduced that can be considered complete isometry invariants in theory.
\smallskip

Now the computational resources are used for generating millions of structures many of which turn out to be near-duplicates.
Problem~\ref{pro:classification} has become the important scientific barrier for paper `milling' by validating any newly discovered crystals vs all known ones. 
The Crystal Isometry Principle and underlying invariants were used for 
mapping the CSD \cite{widdowson2024continuous}, property predictions in \cite{ropers2022fast,balasingham2024accelerating,balasingham2024material} and presented at the IUCr congresses 2021 and 2023, the European Crystallographic Meeting 2022, the BCA annual meetings 2022-2024, MACSMIN 2021-2023 (Mathematics and Computer Science for Materials Innovation).
\smallskip
 
The second author thanks Matt McDermott for a comprehensive tour of the A-lab in Berkeley, where 43 materials were synthesized \cite{szymanski2023autonomous}, 
Ram Seshadri (UCSB), Leslie Schoop (Princeton), and Robert Palgrave (UCL) for
fruitful discussions. 
We thank Andy Cooper (Liverpool), Simon Billinge (Columbia), and all members of the Data Science Theory and Applications group in the Materials Innovation Factory at Liverpool, especially to Daniel Widdowson and Yury Elkin.
\smallskip

This work was supported by the second author's Royal Academy of Engineering Fellowship `Data Science for Next Generation Engineering of Solid Crystalline Materials' (IF2122/186), the EPSRC New Horizons grant `Inverse design of periodic crystals' (EP/X018474/1), and the Royal Society APEX fellowship `New geometric methods for mapping the space of periodic crystals' (APX/R1/231152).
We thank all reviewers for their valuable time and helpful advice.

\referencelist

\end{document}